\documentclass{article}
\usepackage{spconf,graphicx,amsmath,amssymb}
\usepackage{booktabs}   
\usepackage{multirow}   
\usepackage{url}
\usepackage[bottom]{footmisc}
\usepackage{subcaption}

\title{SELF-SUPERVISED RESTORATION OF SINGING VOICE DEGRADED BY PITCH SHIFTING USING SHALLOW DIFFUSION}

\name{Yunyi Liu$^{1}$\thanks{This work was conducted at Sony Computer Science Laboratories, Tokyo, Japan.}
\quad Taketo Akama$^{2}$}

\address{$^{1}$University of Sydney, Electrical and Computer Engineering \\
         $^{2}$Sony Computer Science Laboratories, Tokyo, Japan}

\begin{document}
\maketitle

\begin{abstract}
Pitch shifting has been an essential feature in singing voice production. However, conventional signal processing approaches exhibit well–known trade–offs such as formant shifts and robotic coloration that becomes more severe at larger transposition jumps. This paper targets high–quality pitch shifting for singing by reframing it as a restoration problem: given an audio track that has been pitch–shifted (and thus contaminated by artifacts), we recover a natural–sounding performance while preserving its melody and timing. Specifically, we use a lightweight, mel–space diffusion model driven by frame–level acoustic features such as $F_0$, volume, and content features. We construct training pairs in a self-supervised manner by applying pitch shifts and reversing them to simulate realistic artifacts while retaining ground truth. On a curated singing set, the proposed approach substantially reduces pitch–shift artifacts compared to representative classical baselines, as measured by both statistical metrics and pairwise acoustic measures. The results suggest that restoration–based pitch shifting could be a viable approach towards artifact–resistant transposition in vocal production workflows.
\end{abstract}

\begin{keywords}
singing voice, restoration, diffusion models, vocoder, pitch shifting.
\end{keywords}

\begin{figure*}[ht]
  \centering
  \includegraphics[width=\textwidth]{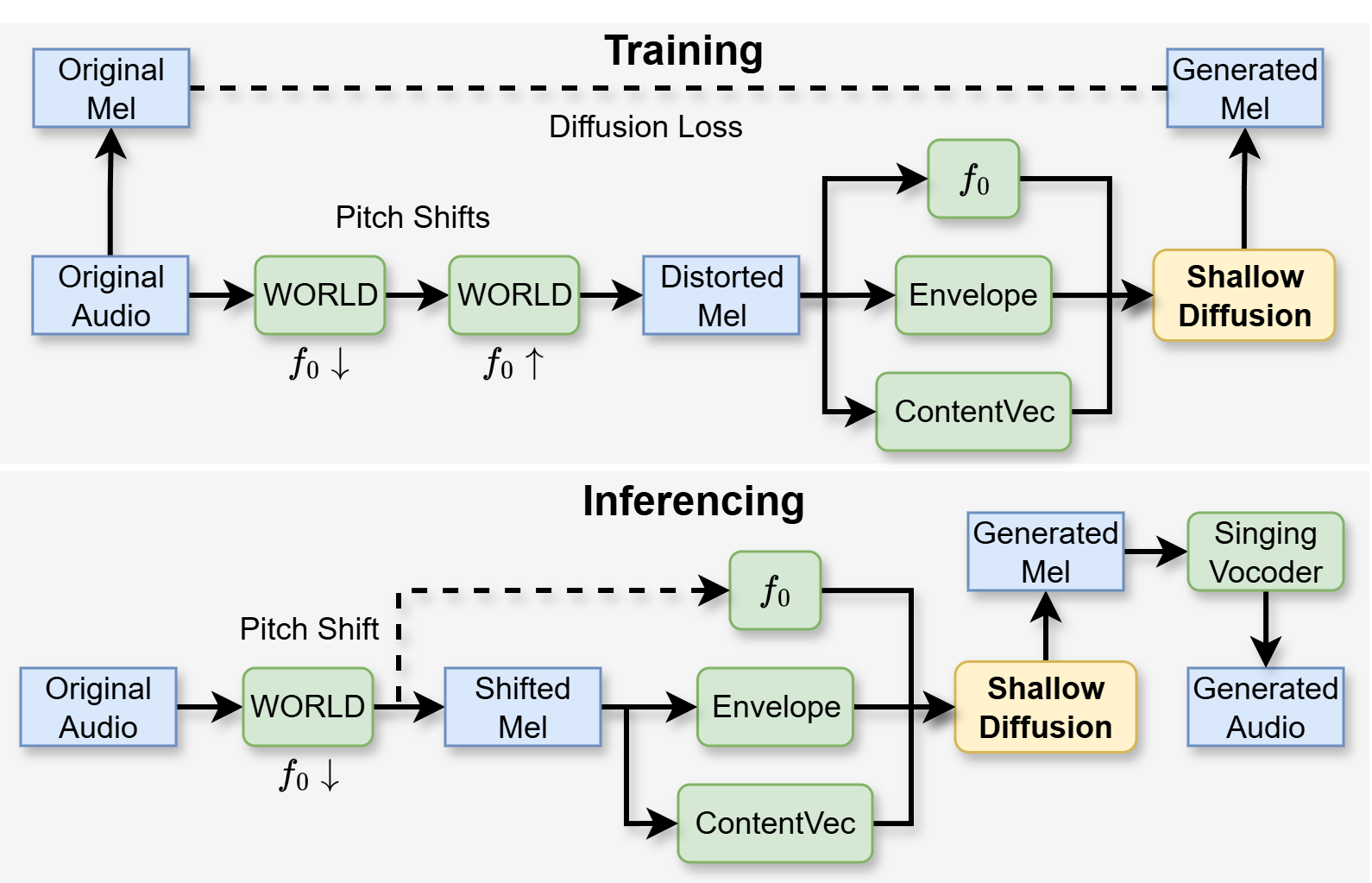}
  \caption{Overview. \textbf{Training}: Original audio goes through a forward and backward pitch shift process via the WORLD vocoder. The shallow diffusion then denoises the artifact audio to reconstruct the mel. Notice $f_0$ is extracted from artifact Mel during training for higher robustness. \textbf{Inference}: Given a shifted audio, the diffusion model reconstructs it into a clean Mel-spectrogram. The $f0$ is directly input to the diffusion model during inference.}
  \label{fig:pipeline}
\end{figure*}

\section{Introduction}

\subsection{Classical Pitch Shift}

Classical pitch shifting algorithms manipulate either the waveform in time or its short-time spectrum. Resampling technique for example, shifts the pitch by scaling time \(t \mapsto t/r\) (or samples \(n \mapsto n/r\)) with \(ratio = 2^{\Delta/12}\), and then uses a separate time–scale modification stage to restore the original duration. This simple pipeline drags the spectral envelope (formants) and can misalign transients, which could yield chipmunk colouration unless additional formant or envelope correction is applied. 
\\
\\
Phase–vocoder methods analyze frames with an STFT, unwrap phases to estimate instantaneous frequency, and then resynthesize the signal using modified phase increments corresponding to the desired pitch ratio. With peak phase locking and transient protection~\cite{Laroche1999PhaseVocoder,Roebel2010PhaseLock}, they preserve the broad spectral shape, but large pitch shifts or rapid modulations can still produce phasiness effects and attack smearing. Time-domain overlap-and-add (TDOLA) methods avoid using explicit Fourier manipulation. For example, pitch-synchronous overlap-and-add (PSOLA) detects pitch marks and extracts pitch-synchronous grains windowed around each mark, and overlap-adds them at a new period to shift pitch. On the other hand, waveform-similarity OLA~\cite{Verhelst1993WSOLA,Driedger2014WSOLADetail} uses fixed-hop grains aligned by short-time cross-correlation before overlap-add. These approaches offer low latency and crisp transients for small ratios on voiced material, but they are sensitive to alignment errors and tend to drift formants without extra correction, and can sound buzzy in unvoiced or noisy regions. 
\\
\\
Analysis–synthesis vocoders~\cite{Morise2016WORLD, Kawahara2001STRAIGHT} firstly estimate important acoustic features such as $f_0$, spectral envelope, and band aperiodicity. It then resynthesizes the signal using harmonic excitation plus filtered noise. This provides explicit pitch and formant control and can preserve timbre under moderate shifts. However, when globally scaling the entire \(F_{0}\) contour, the envelope or aperiodicity estimation errors and a single-excitation model often lead to artifacts such as robotic colorization. 
\\
\\
Therefore, although classical signal-processing methods provide controllable pitch manipulation, they are generally prone to audible artifacts unless carefully tuned, and their degradations typically take place as the magnitude of the pitch transposition increases.

\subsection{Data-driven Approaches}

Data-driven approaches instead learn priors over natural singing spectra from large corpora, which enables more robust timbre preservation. Modern singing voice synthesis/conversion (SVS/SVC) systems typically map symbolic or self-supervised content features~\cite{Hubert,contentvec} to mel or waveform using neural decoders and high-fidelity vocoders. This formulation allows pitch editing by directly modifying the frame-level \(f_{0}\) contour while relying on the model’s learned prior to maintain vocal identity.

Several recent models incorporate explicit mechanisms for pitch conditioning. Diff-Pitcher~\cite{diff-pitcher} applies learned pitch perturbation inside a diffusion prior to improve timbre stability under pitch shifts. SiFiGAN~\cite{Sifigan} conditions a sine-excitation generator on \(f_{0}\) for harmonic consistency, while NeuroDyne~\cite{neurodyne} adopts a neural source–filter structure to provide interpretable control over excitation and formants.

Diffusion models~\cite{DDPM} have become popular in SVS/SVC pipelines due to their strong generative priors. Early vocoders such as DiffWave and WaveGrad~\cite{Kong2021DiffWave,Chen2021WaveGrad} required hundreds of sampling steps, but later systems such as diffSinger~\cite{Liu2022DiffSinger} and Grad-SVC~\cite{GradSVC} improved pitch stability and sampling efficiency through conditioning on score, \(f_{0}\), and content features. Sampling has been further accelerated by DDIM~\cite{Song2021DDIM}, DPM-Solver~\cite{Lu2022DPMSolver}, distillation~\cite{Salimans2022ProgressiveDistillation}, and consistency decoding~\cite{Song2023Consistency}. Other hybrid methods, such as DDSP-SVC~\cite{DDSP-SVC}, combine diffusion with differentiable source–filter synthesis, while LDM-SVC~\cite{Chen2024LDM-SVC} performs conversion in a latent diffusion space for near–zero-shot generalization.

However, most diffusion-based SVS/SVC systems rely on speaker embeddings and thus are not directly applicable to source-agnostic pitch shifting of unseen singers. Motivated by this limitation, shallow diffusion~\cite{Liu2022DiffSinger, DDSP-SVC} uses a diffusion model as a lightweight refiner initialized from a strong prior which requires only tens of denoising steps. Our work extends shallow diffusion by leveraging it within a novel architecture that incorporates shallow diffusion specifically for noise reduction after a WORLD-based pitch shift. While shallow diffusion remains at the core of the noise reduction process, our work introduces an architecture that utilizes shallow diffusion in a more generalized framework which aims at improving the pitch-shifting process for source-agnostic scenarios. At inference, the restored mel is rendered to waveform using an NSF-HiFiGAN-style vocoder~\cite{Zheng2023NSFHiFiGAN}.

\section{Method}

\begin{table*}[t]
\centering
\caption{Summary of singing voice datasets used in this study. The combined dataset provides diverse coverage across languages, singing styles, and vocal characteristics.}
\label{tab:datasets}
\resizebox{\textwidth}{!}{%
\begin{tabular}{@{}llllll@{}}
\toprule
\textbf{Dataset} & \textbf{Language(s)} & \textbf{Singers} & \textbf{Duration} & \textbf{Content} & \textbf{Key Features} \\ \midrule
CSD & Korean, English & 1 (F) & $\sim$200 min & Children's songs & MIDI, lyrics, F3-F5 \\
NUS-48E & English & 12 (M/F) & 169 min & Pop songs & Phone annotations \\
VocalSet & English & 20 (9M, 11F) & 606 min & Vocal techniques & Extended techniques \\
Choral & Latin, Spanish, Catalan & 16 (SATB) & $\sim$15 min & A cappella choir & Multi-track, F0 \\
ESMUC & German, Latin & 12 (SATB) & 31 min & Choral pieces & F0, note annotations \\
\bottomrule
\end{tabular}%
}
\end{table*}

\subsection{Data Pairs Generation}

One of the main challenges for data-driven pitch shift models is the lack of ground-truth pitch shift pairs. As conventional methods introduce artifacts inevitably after performing pitch transposition, it is difficult to obtain perfect ground-truth singing voice sounds that are shifted from a singing voice. Therefore, our self-supervised strategy for this issue is to degrade the audio first as inputs and use the original audio as ground-truth. As shown in Figure~\ref{fig:pipeline}, we firstly use WORLD vocoder to perform pitch shifts up on a singing sound (by certain semitones between -12 and 12 semitones). Then we use the same WORLD vocoder to pitch shift the audio back under the same configuration. The result of this is a distorted singing sound that is supposed to be of same pitch, volume, and speaker characteristics due to the imperfect reconstruction of WORLD vocoder. We use WORLD as the prior because its source–filter structure gives exact pitch tracking and strong zero-shot robustness, unlike neural vocoders that may drift when encountering unseen singers. WORLD sets a clean pitch contour, and the diffusion model removes artifacts while keeping the pitch intact. To justify our choice, we show why it is necessary to include WORLD for $f_0$ conditioning in Section~\ref{sec: pitch-accuracy}.

For the distorted audio, we extract acoustic features following a similar approach to DiffWaveNetSVC~\cite{DiffWaveNetSVC}. Specifically, we use: (1) fundamental frequency ($f_0$) extracted via the Crepe~\cite{crepe} algorithm, (2) volume envelope computed from frame-level RMS energy, and (3) content features from ContentVec~\cite{contentvec}, which provides speaker-independent phonetic representations. These features were pre-processed to serve as inputs to our diffusion model.

\subsection{Shallow Diffusion Model}

Our model follows a shallow variant of a denoising diffusion probabilistic model (DDPM)~\cite{DDPM} operating directly in the mel-spectrogram domain, and is based on the implementation of Diffusion-SVC~\cite{Diffusion-SVC}. The model is a 1-D temporal U-Net with 20 residual layers, each containing 512 channels and 256 hidden units. We condition the denoising network on pitch ($f_0$), volume, and ContentVec~\cite{contentvec} features via adaptive normalization layers. The diffusion process uses $K=100$ discrete noise steps, and the DPM-Solver sampling method~\cite{Lu2022DPMSolver} with a speedup factor of 10 is employed at inference time. 

The forward process gradually adds Gaussian noise to the clean mel-spectrogram $x_0$ to produce $x_t$:
\[
q(x_t|x_0) = \mathcal{N}(\sqrt{\bar{\alpha}_t}x_0, (1-\bar{\alpha}_t)\mathbf{I}),
\]
while the reverse model predicts the noise $\epsilon_\theta(x_t, c, t)$ conditioned on acoustic features $c$ (pitch, volume, content). 
The denoising objective is the standard $L_2$ diffusion loss:
\[
\mathcal{L}_{\text{diff}} = \mathbb{E}_{x_0, \epsilon, t}\left[\|\epsilon - \epsilon_\theta(x_t, c, t)\|_2^2\right].
\]
To further stabilize pitch and energy restoration, we include auxiliary $L_1$ losses between predicted and reference mel-spectrograms and $f_0$: 

\[
\mathcal{L}_{\mathrm{mel}} = \|\hat{M} - M\|_{1}, 
\qquad
\mathcal{L}_{F_0} = \|\hat{f}_0 - f_0\|_{1}.
\]

The total training objective is thus:
\[
\mathcal{L} = \mathcal{L}_{\text{diff}} + \lambda_1 \mathcal{L}_{\text{mel}} + \lambda_2 \mathcal{L}_{f_0},
\]
where $\lambda_1$ and $\lambda_2$ are empirically set to balance reconstruction and conditioning accuracy.

\subsection{Training Procedure}

We train our model on a diverse mixture of five singing voice datasets: CSD~\footnote{\url{https://program.ismir2020.net/static/lbd/ISMIR2020-LBD-435-poster/index.html}}, NUS-48E\footnote{\url{https://opendatalab.com/OpenDataLab/NUS-48E}}, VocalSet~\footnote{\url{https://zenodo.org/records/1193957}}, Choral Singing Dataset~\footnote{\url{https://zenodo.org/records/1286570}}, and ESMUC~\footnote{\url{https://zenodo.org/records/5848990}}, as shown in Table~\ref{tab:datasets}. These datasets collectively span multiple languages (English, Korean, Chinese, German, Latin, Spanish, and Catalan), various singing styles (pop, folk, classical, choral), and diverse vocal characteristics from 61 unique singers. This heterogeneous mixture ensures robust cross-lingual and cross-style generalization of our pitch shifting restoration model. All audio is resampled to 44.1\,kHz and converted into 128-dimensional log-mel spectrograms ($n_\mathrm{fft}=1024$, hop size $=256$, window size $=1024$, $f_{\min}=0$, $f_{\max}=22{,}050$). We split the corpus into training and validation sets with a ratio of 0.95/0.05. Each training sample corresponds to a pair of distorted and clean mel-spectrogram segments of 2\,s duration (block size 512 frames). 

During preprocessing, the $f_0$ contour is extracted using the CREPE~\cite{crepe} pitch estimator, with limits in 65–800\,Hz. For each segment, we also compute the volume envelope and extract content embeddings using the pretrained ContentVec~\cite{contentvec} encoder, which was trained on 16\,kHz audio, 768-dimensional output with a hop size 320. These features are concatenated and used as conditioning for the diffusion model, while the target output is the clean mel-spectrogram segment.

We trained the model for 10,000 epochs with a batch size of 32 using the Adam optimizer (learning rate $2\times10^{-4}$, weight decay 0) on an NVIDIA RTX 3090 GPU. The learning rate decays by a factor $\gamma=0.5$ every 10{,}000 steps. We also used automatic mixed precision to enable training stability. The validation is performed every 2,000 steps.

\subsection{Inference}

During the usage stage, we use the same WORLD vocoder to perform pitch shift, which may introduce artifacts. Let \(c=[\,u,\ f_0,\ e\,]\) denote conditioning features (content units, pitch, and energy) extracted from the artifact audio, and let \(M_{\mathrm{art}}\) be a Mel spectrogram computed from the same artifact.
We initialize the latent at step \(T\) by injecting noise to the artifact Mel spectrogram:
\[
x_T \;=\; \sqrt{\bar{\alpha}_T}\,M_{\mathrm{art}} \;+\; \sqrt{1-\bar{\alpha}_T}\,z, 
\qquad z \sim \mathcal{N}(0,\mathbf{I}).
\]
At each step \(t\) the network predicts noise \(\epsilon_\theta(x_t,c,t)\), from which we form the clean estimate
\[
\hat{x}_0(x_t,c,t) \;=\; \frac{x_t - \sqrt{1-\bar{\alpha}_t}\,\epsilon_\theta(x_t,c,t)}{\sqrt{\bar{\alpha}_t}}
\]
Using a deterministic DDIM/DPM-style update (\(\eta{=}0\)), we step:
\begin{equation*}
\begin{aligned}
x_{t-1}
&= \sqrt{\bar{\alpha}_{t-1}}\,\hat{x}_0(x_t,c,t)
  + \sqrt{1-\bar{\alpha}_{t-1}}\,\epsilon_\theta(x_t,c,t),\\
t &\in \{T,T\!-\!1,\dots,1\}.
\end{aligned}
\end{equation*}
After \(K\) shallow steps, we take \(\hat{M}=x_0\) as the restored mel-spectrogram and synthesize waveform with an \(F_0\)-aware pretrained NSF-HiFiGAN vocoder~\cite{Zheng2023NSFHiFiGAN}.

\begin{table*}[!htb]
\centering
\setlength{\tabcolsep}{4pt}
\renewcommand{\arraystretch}{0.95}
\resizebox{\textwidth}{!}{%
  \begin{tabular}{lrrrrrrrrr}
\toprule
& \multicolumn{3}{c}{\textbf{Statistical}} & \multicolumn{6}{c}{\textbf{Pairwise}} \\
\cmidrule(lr){2-4}\cmidrule(lr){5-10}
\textbf{System} & \textbf{FAD$\downarrow$} & \textbf{KID$\downarrow$} & \textbf{MMD$\downarrow$} &
\textbf{SC$\downarrow$} & \textbf{LSD$\downarrow$} & \textbf{SI-SDR$\downarrow$} &
\textbf{F0$\downarrow$} & \textbf{V/UV$\downarrow$} & \textbf{MFCC$\downarrow$} \\
\midrule
PSOLA            & 19.248 & 0.0083 & 0.1607 & 0.8442 & \textbf{0.6434} & -33.407 & 41.776 & 6.06\% & 39.059 \\
WORLD   & 31.397 & 0.0271 & 0.2287 & 1.4766 & 3.573  & -47.836 & 36.555 & 13.26\% & 97.625 \\
CLPCNet          & 77.702 & 0.0861 & 0.3644 & 0.9067 & 3.3673 & \textbf{-48.504} & 16.946 & 24.12\% & 65.323 \\
SiFiGAN & 49.048 & 0.0792 & 0.4382 & 0.5526 & 3.0687 & -34.073 & 20.608 & 9.43\% & 54.754 \\
Diff-Pitcher & 39.321 & 0.0333 & 0.2740 & 0.6366 & 2.7519 & -42.938 & 27.439 & 14.07\% & 37.775 \\
\textbf{Ours}    & \textbf{7.886} & \textbf{0.0030} & \textbf{0.0921} & \textbf{0.0290} & 0.7811 & -33.636 & \textbf{2.881} & \textbf{0.33\%} & \textbf{23.616} \\
\bottomrule
\end{tabular}%
}
\caption{Objective evaluation results of pitch-shifted sounds quality and similarity relative to their unshifted versions.}
\label{tab: results}
\end{table*}

\section{Evaluation}

\subsection{Dataset}
We evaluate on the \emph{Singing Voice Audio Dataset}~\footnote{\url{https://zenodo.org/records/3534236}}, which is unseen from training. It contains over 70 unaccompanied vocal recordings by 28 professional, semi-professional, and amateur singers. Most excerpts are Chinese Opera with some Western Opera, sampled at 44.1kHz and amplitude-normalized. This choice targets the analysis of singing voice without accompaniment and ensures no speaker or content overlap with the training dataset.

\subsection{Comparison}
As our task is high-quality pitch transposition via restoration, for the proposed shallow-diffusion model, we feed the artifact audio and synthesize the restored signal. We select five variants of pitch shifting algorithms for comparison: time-domain pitch-synchronous overlap-and-add (TD-PSOLA)~\cite{Moulines1990PSOLA}~\footnote{\url{https://github.com/maxrmorrison/psola}}, Controllable LPCNET~\cite{clpcnet}\footnote{\url{https://github.com/maxrmorrison/clpcnet?tab=readme-ov-file}}, WORLD\cite{Morise2016WORLD}, SiFiGAN~\cite{Sifigan}, and Diff-Pitcher~\cite{diff-pitcher}. We run the same forward–reverse pitch shift procedure and resynthesize with each method’s standard settings, and then compare every generated file against its corresponding clean reference. All audios are evaluated at 44.1kHz with a hop size of 512 samples. 

\subsection{Metrics}

We evaluate our results on both statistical distribution metrics and pairwise signal difference as shown in Table~\ref{tab: results}. The former measures global distribution similarity between generated and real audio, while the latter evaluates clip-level reconstruction fidelity relative to clean references. \textit{Frech\'et Audio Distance (FAD)}~\cite{Kilgour2019FAD} measures the distribution gap between VGGish embeddings of restored and reference audio using the standard 10\,s window protocol. \textit{Kernel Inception Distance (KID)}~\cite{KID} quantifies the squared MMD between generated and reference embedding distributions using polynomial kernels. Lower KID means better distribution matching. We use KID because it is unbiased and more stable for smaller datasets. Finally, \textit{Maximum Mean Discrepancy (MMD)}~\cite{MMD} measures the mean embedding difference between the generated and reference distributions. We use an RBF kernel over VGGish features to compute the exact two-sample divergence.

For pairwise comparisons, we employ a combination of metrics. \textit{Spectral Convergence (SC)} evaluates how well strong spectral components are preserved. \textit{Log-Spectral Distance (LSD)} quantifies perceptual harmonic distortion. \textit{Scale-Invariant SDR (SI-SDR)} measures waveform fidelity while ignoring amplitude mismatch. Higher SI-SDR indicates fewer temporal and phase artifacts. \textit{Pitch Error (F0)} evaluates the measured fundamental frequency $f_0$ from CREPE in RMSE in cents, which reflects harmonic stability recovery after WORLD distortion. \textit{Voiced/Unvoiced Error Rate (V/UV)} measures the binary mismatch of voiced and unvoiced states. This penalizes missing or false pitch activation which are crucial to singing articulation. Finally, \textit{MFCC} computes the mean $L_{2}$ distance between 13-dimensional MFCCs which captures the timbre and spectral envelope distortions.

\begin{table*}[t]
\centering
\small
\setlength{\tabcolsep}{4.5pt}
\renewcommand{\arraystretch}{1.25}

\begin{tabular}{l|
cc|cc|cc|cc|cc|cc}
\hline
\multirow{2}{*}{\textbf{Method}} &
\multicolumn{2}{c|}{\textbf{$-12$}} &
\multicolumn{2}{c|}{\textbf{$-6$}} &
\multicolumn{2}{c|}{\textbf{$-3$}} &
\multicolumn{2}{c|}{\textbf{$+3$}} &
\multicolumn{2}{c|}{\textbf{$+6$}} &
\multicolumn{2}{c}{\textbf{$+12$}} \\
 & $f_{0}\downarrow$ & V/UV\%$\downarrow$
 & $f_{0}\downarrow$ & V/UV\%$\downarrow$
 & $f_{0}\downarrow$ & V/UV\%$\downarrow$
 & $f_{0}\downarrow$ & V/UV\%$\downarrow$
 & $f_{0}\downarrow$ & V/UV\%$\downarrow$
 & $f_{0}\downarrow$ & V/UV\%$\downarrow$ \\
\hline

\textbf{Diff-Pitcher} &
0.071 & 5.421 &
0.036 & 5.740 &
0.028 & 5.700 &
0.030 & 6.080 &
0.052 & 7.840 &
0.210 & 10.66 \\

\textbf{SiFiGAN} &
0.067 & 8.870 &
0.064 & 7.430 &
0.057 & 6.960 &
0.053 & 6.120 &
0.065 & 8.400 &
0.142 & 14.23 \\

\textbf{TD-PSOLA} &
0.156 & 0.368 &
2.222 & 13.55 &
2.076 & 6.971 &
2.105 & 7.494 &
2.313 & 12.52 &
2.563 & 12.86 \\

\textbf{CLPCNet} &
2.246 & 11.70 &
1.896 & 8.890 &
1.548 & 7.270 &
1.411 & 6.411 &
1.647 & 7.303 &
1.736 & 8.030 \\

\textbf{WORLD} &
0.009 & 6.006 &
0.008 & 4.173 &
0.010 & 3.723 &
0.023 & 3.269 &
0.018 & 4.130 &
0.205 & 4.755 \\

\textbf{Diff-only} &
0.694 & 4.353 &
0.371 & 8.927 &
0.014 & 6.318 &
0.100 & 15.82 &
0.384 & 55.58 &
0.428 & 24.52 \\

\textbf{Ours} &
0.012 & 6.145 &
0.009 & 4.465 &
0.010 & 4.219 &
0.012 & 4.252 &
0.013 & 4.317 &
0.188 & 7.112 \\

\hline

\end{tabular}

\caption{Pitch shift accuracy results showing the RMSE of log $f_0$ and voiced/unvoiced decision error rate (V/UV) for pitch transposition. Lower values are better.}
\label{tab:f0}
\end{table*}

\section{Results}
\label{sec:results}

\subsection{Audio Reconstruction}

Our shallow diffusion system, denoted as ‘Ours’, achieves the best statistical comparison performance. As shown in Table~\ref{tab: results}, our shallow diffusion model produces outputs that are both distributionally closest to the clean corpus and spectrally most faithful to the references. We obtain the lowest FAD, KID, and MMD scores, outperforming the other approaches by a large margin. This indicates that the global embedding distribution of our reconstructions is closest to real singing and aligns with the model’s role as a mel-space denoiser for pitch-shift artifacts.

Our method also achieves the best results on spectral convergence (0.0290), MFCC (23.616), F0 RMSE (2.881 cents), and voicing error (0.33\%). The strong $F_0$ and V/UV scores are expected: we condition the diffusion model directly on ground-truth $f_{0}$, volume, and content features, which stabilize the overall pitch contour and spectral distributions. This allows the diffusion model to focus on removing pitch-coupled artifacts rather than re-estimating pitch or content. TD-PSOLA attains the best LSD (0.6434), and CLPCNet achieves the highest SI--SDR ($-48.504$\,dB). In practice, LSD is computed on log STFT magnitudes and is sensitive to narrowband harmonics. TD-PSOLA can retain fine-bin spectral detail in steady voiced regions even when timbre degrades, which explains its advantage despite weaker mel/MFCC/F0/V/UV results. CLPCNet's SI-SDR advantage stems from retaining more temporal fine structure through its LPC-based excitation, though this does not translate to perceptual or mel-spectral fidelity. The large gap in MFCC relative to TD-PSOLA and CLPCNet reflects more accurate spectral envelopes after denoising and vocoding. Our method remains competitive on SI--SDR ($-33.636$\,dB) while leading in the most perceptually aligned metrics.

SiFiGAN and Diff-Pitcher have show mixed performance in this reconstruction setting. Both models were evaluated in a zero-shot manner and they were not trained on our dataset or singers. Their spectral statistics and pairwise metrics strongly depend on how well their pretrained priors generalize. SiFiGAN achieves moderate performance on LSD (3.0687) and SC (0.5526), but its $F_0$ error (20.608 cents) and MFCC L2 (54.754) reveal limited robustness for unseen voices, especially male singers whose low-$f_{0}$ regions challenge its sinusoidal synthesizer. Diff-Pitcher performs slightly better than SiFiGAN on MFCC and SC, but remains behind our method in FAD, F0 accuracy, and timbral reconstruction, likely because its diffusion-based pitch transposition module is optimized for pitch editing rather than artifact-free reconstruction.

\subsection{Pitch Accuracy}
\label{sec: pitch-accuracy}

We acknowledge that in real-world applications, pitch shift is typically performed in one direction, either upward or downward. Therefore to better simulate this setting, we also report the root mean square error of log $f_0$ and voiced/unvoiced decision error rate. In Table~\ref{tab:f0}, we show the results by performing pitch shifts using the same models on the same test dataset. In addition, we also report the result of inferencing our shallow-diffusion model alone denoted as 'Diff-only' by inputting the shifted $f_0$ as well as the volume envelope and content features extracted from the original audio as an ablation study. 

Among all systems, \textit{WORLD} exhibits the strongest baseline pitch stability across a wide range of semitone shifts. This is expected given its explicit source-filter structure, where the excitation is directly parameterized by the target $f_0$ contour. Because \textit{WORLD} deterministically enforces the specified $f_0$, it provides an accurate pitch trajectory even under large transpositions. However, while WORLD preserves pitch well, its formant shifting behaviour introduces audible artifacts. As our method combines \textit{WORLD} with a shallow diffusion denoiser, we preserve \textit{WORLD}'s reliable pitch tracking while reducing spectral artifacts introduced by the source-filter resynthesis. Since the diffusion model only performs a small number of refinement steps on a WORLD-generated mel-spectrogram, it does not modify the underlying harmonic trajectory. As a result, our maintains close pitch accuracy while providing substantially improved timbre quality and artifact removal, as observed in the previous section.

In contrast, \textit{SiFiGAN}, \textit{Diff-Pitcher}, and \textit{CLPCNet} operate in a zero-shot inference setting. In consequence, their ability to follow extreme pitch manipulations depends on how well their pretrained models generalize. Diff-Pitcher, while designed for controllable pitch perturbation, still exhibits increasing pitch drift at higher transposition magnitudes. SiFiGAN, which relies on a sine-based excitation prior, shows even larger deviations. This was a known limitation especially for male voices as the model was pre-trained on female singers. \textit{CLPCNet}, as a neural LPC-based vocoder, relies on autoregressive excitation generation and therefore suffers from pitch inaccuracies when encountering unseen speakers or large pitch shifts. 

Overall, these findings show that WORLD is necessary for reliable fundamental frequency control, and that a shallow diffusion refiner can correct the artifacts introduced by WORLD without compromising its pitch fidelity. This makes our approach particularly suitable for source-agnostic pitch shifting, where consistent pitch accuracy must be retained even when shifting unseen singers.

\section{Conclusion}
In this paper, we presented a simple yet effective singing voice restoration system that learns to undo pitch-shift artifacts. The method is built on top of a classical vocoder by introducing an audio restoration model that utilizes shallow diffusion in the Mel space. On an unseen singing voice audio dataset, the method excels on both distribution and pairwise metrics. We have also shown why it is necessary to incorporate the World vocoder as part of our architecture, as it provides crucial acoustic priors and stabilitizes the pitch so that our diffusion model could focus entirely on audio reconstruction and enhancement. 

\clearpage
\newpage

\begingroup
\renewcommand{\footnotesize}{\normalsize}
\bibliographystyle{IEEEtran}
\bibliography{refs}

@article{Laroche1999PhaseVocoder,
  author    = {Jean Laroche and Mark Dolson},
  title     = {Improved Phase Vocoder Time-Scale Modification of Audio},
  journal   = {IEEE Transactions on Speech and Audio Processing},
  volume    = {7},
  number    = {3},
  pages     = {323--332},
  year      = {1999}
}

@inproceedings{Roebel2010PhaseLock,
  TITLE = {{A new approach to transient processing in the phase vocoder}},
  AUTHOR = {Roebel, Axel},
  URL = {https://hal.science/hal-01161124},
  NOTE = {                            cote interne IRCAM: Roebel03a},
  BOOKTITLE = {{6th International Conference on Digital Audio Effects (DAFx)}},
  ADDRESS = {London, United Kingdom},
  PAGES = {344-349},
  YEAR = {2003},
  MONTH = Sep,
  HAL_ID = {hal-01161124},
  HAL_VERSION = {v1},
}

@inproceedings{Kilgour2019FAD,
  title     = {Fréchet Audio Distance: A Reference-Free Metric for Evaluating Music Enhancement Algorithms},
  author    = {Kevin Kilgour and Mauricio Zuluaga and Dominik Roblek and Matthew Sharifi},
  booktitle = {Proceedings of Interspeech 2019},
  pages     = {2350-2354},
  year      = {2019},
  url       = {https://www.isca-archive.org/interspeech_2019/kilgour19_interspeech.html}
}

@misc{clpcnet,
      title={Neural Pitch-Shifting and Time-Stretching with Controllable LPCNet}, 
      author={Max Morrison and Zeyu Jin and Nicholas J. Bryan and Juan-Pablo Caceres and Bryan Pardo},
      year={2021},
      eprint={2110.02360},
      archivePrefix={arXiv},
      primaryClass={eess.AS},
      url={https://arxiv.org/abs/2110.02360}, 
}

@article{Moulines1990PSOLA,
title = {Pitch-synchronous waveform processing techniques for text-to-speech synthesis using diphones},
journal = {Speech Communication},
volume = {9},
number = {5},
pages = {453-467},
year = {1990},
note = {Neuropeech '89},
issn = {0167-6393},
doi = {https://doi.org/10.1016/0167-6393(90)90021-Z},
author = {Eric Moulines and Francis Charpentier},
}

@inproceedings{Verhelst1993WSOLA,
author = {Verhelst, Werner and Roelands, Marc},
title = {An overlap-add technique based on waveform similarity (WSOLA) for high quality time-scale modification of speech},
year = {1993},
isbn = {0780309464},
publisher = {IEEE Computer Society},
address = {USA},
booktitle = {Proceedings of the 1993 IEEE International Conference on Acoustics, Speech, and Signal Processing: Speech Processing - Volume II},
pages = {554–557},
numpages = {4},
location = {Minneapolis, Minnesota, USA},
series = {ICASSP'93}
}

@inproceedings{Driedger2014WSOLADetail,
  title={TSM Toolbox: MATLAB Implementations of Time-Scale Modification Algorithms},
  author={Jonathan Driedger and Meinard M{\"u}ller},
  booktitle={International Conference on Digital Audio Effects},
  year={2014},
}

@article{Kawahara2001STRAIGHT,
  author    = {Hideki Kawahara and Ikuyo Masuda-Katsuse and Alain de Cheveign{\'e}},
  title     = {Restructuring Speech Representations Using a Pitch-Adaptive Time--Frequency Smoothing and an Instantaneous-Frequency-Based {F0} Extraction: {STRAIGHT}},
  journal   = {Speech Communication},
  volume    = {10},
  number    = {1},
  pages     = {55--63},
  year      = {2001}
}

@article{Morise2016WORLD,
  author    = {Masashi Morise and Fumiya Yokomori and Kenji Ozawa},
  title     = {{WORLD}: A Vocoder-Based High-Quality Speech Synthesis System for Real-Time Applications},
  journal   = {IEICE Transactions on Information and Systems},
  volume    = {E99.D},
  number    = {7},
  pages     = {1877--1884},
  year      = {2016}
}

@inproceedings{Liu2022DiffSinger,
  title     = {DiffSinger: Singing Voice Synthesis via Shallow Diffusion Mechanism},
  author    = {Liu, Jinglin and Li, Chengxi and Ren, Yi and Chen, Feiyang and Zhao, Zhou},
  booktitle = {Proceedings of the AAAI Conference on Artificial Intelligence},
  volume    = {36},
  number    = {10},
  pages     = {11020--11028},
  year      = {2022},
  doi       = {10.1609/aaai.v36i10.21350},
}

@inproceedings{Kong2021DiffWave,
  title     = {DiffWave: A Versatile Diffusion Model for Audio Synthesis},
  author    = {Kong, Zhifeng and Ping, Wei and Huang, Jiaji and Zhao, Kexin and Catanzaro, Bryan},
  booktitle = {International Conference on Learning Representations (ICLR)},
  year      = {2021},
}

@inproceedings{Chen2021WaveGrad,
  title     = {WaveGrad: Estimating Gradients for Waveform Generation},
  author    = {Chen, Nanxin and Zhang, Yu and Zen, Heiga and Weiss, Ron J. and Norouzi, Mohammad and Chan, William},
  booktitle = {International Conference on Learning Representations (ICLR)},
  year      = {2021},
}

@inproceedings{Song2021DDIM,
  title     = {Denoising Diffusion Implicit Models},
  author    = {Song, Jiaming and Meng, Chenlin and Ermon, Stefano},
  booktitle = {International Conference on Learning Representations (ICLR)},
  year      = {2021},
}

@inproceedings{Lu2022DPMSolver,
  title     = {DPM-Solver: A Fast {ODE} Solver for Diffusion Probabilistic Model Sampling in Around 10 Steps},
  author    = {Lu, Cheng and Zhou, Yuhao and Bao, Fan and Chen, Jianfei and Li, Chongxuan and Zhu, Jun},
  booktitle = {Advances in Neural Information Processing Systems (NeurIPS)},
  year      = {2022},
}

@inproceedings{Salimans2022ProgressiveDistillation,
  title     = {Progressive Distillation for Fast Sampling of Diffusion Models},
  author    = {Salimans, Tim and Ho, Jonathan},
  booktitle = {International Conference on Learning Representations (ICLR)},
  year      = {2022},
}

@inproceedings{Song2023Consistency,
  title     = {Consistency Models},
  author    = {Song, Yang and Dhariwal, Prafulla and Chen, Mark and Sutskever, Ilya},
  booktitle = {Proceedings of the 40th International Conference on Machine Learning},
  series    = {Proceedings of Machine Learning Research},
  volume    = {202},
  pages     = {32211--32252},
  year      = {2023},
  publisher = {PMLR},
}

@INPROCEEDINGS{diff-pitcher,
  author={Hai, Jiarui and Elhilali, Mounya},
  booktitle={2023 IEEE Workshop on Applications of Signal Processing to Audio and Acoustics (WASPAA)}, 
  title={Diff-Pitcher: Diffusion-Based Singing Voice Pitch Correction}, 
  year={2023},
  pages={1-5},
  doi={10.1109/WASPAA58266.2023.10248127}}

@INPROCEEDINGS{Sifigan,
  author={Yoneyama, Reo and Wu, Yi-Chiao and Toda, Tomoki},
  booktitle={ICASSP 2023 - 2023 IEEE International Conference on Acoustics, Speech and Signal Processing (ICASSP)}, 
  title={Source-Filter HiFi-GAN: Fast and Pitch Controllable High-Fidelity Neural Vocoder}, 
  year={2023},
  pages={1-5},
  doi={10.1109/ICASSP49357.2023.10095298}}

@misc{neurodyne,
      title={Neurodyne: Neural Pitch Manipulation with Representation Learning and Cycle-Consistency GAN}, 
      author={Yicheng Gu and Chaoren Wang and Zhizheng Wu and Lauri Juvela},
      year={2025},
      eprint={2505.15368},
      archivePrefix={arXiv},
      primaryClass={cs.SD},
      url={https://arxiv.org/abs/2505.15368}, 
}

@inproceedings{Zheng2023NSFHiFiGAN,
  author    = {Wenjie Zheng and Yi Ren and Jinglin Liu and et al.},
  title     = {{NSF}-{HiFiGAN}: A Source-Filter Architecture for High-Fidelity Speech Synthesis},
  booktitle = {Proceedings of Interspeech},
  year      = {2023}
}

@inproceedings{GradSVC,
  title={Grad-TTS: A Diffusion Probabilistic Model for Text-to-Speech},
  author={Vadim Popov and Ivan Vovk and Vladimir Gogoryan and Tasnima Sadekova and Mikhail Kudinov},
  booktitle={International Conference on Machine Learning},
  year={2021},
  url={https://api.semanticscholar.org/CorpusID:234483016}
}

@misc{DDSP-SVC,
  title        = {DDSP-SVC: Singing Voice Conversion with Differentiable DSP},
  howpublished = {\url{https://github.com/yxlllc/DDSP-SVC}},
  year         = {2024},
  note         = {Github repository}
}

@misc{Diffusion-SVC,
  title        = {Diffusion-SVC: Diffusion for Singing Voice Conversion},
  howpublished = {\url{https://github.com/CNChTu/Diffusion-SVC}},
  year         = {2023},
  note         = {Github repository}
}

@inproceedings{contentvec,
  title        = {ContentVec: An Improved Self-Supervised Speech Representation by Disentangling Speakers},
  author       = {Kaizhi Qian and Yang Zhang and Heting Gao and Junrui Ni and Cheng-I Lai and David Cox and Mark Hasegawa-Johnson and Shiyu Chang},
  booktitle    = {Proceedings of the 39th International Conference on Machine Learning (ICML)},
  volume       = {162},
  pages        = {18003--18017},
  year         = {2022},
  publisher    = {Proceedings of Machine Learning Research (PMLR)},
  url          = {https://proceedings.mlr.press/v162/qian22b.html}
}

@article{Hubert,
author = {Hsu, Wei-Ning and Bolte, Benjamin and Tsai, Yao-Hung Hubert and Lakhotia, Kushal and Salakhutdinov, Ruslan and Mohamed, Abdelrahman},
title = {HuBERT: Self-Supervised Speech Representation Learning by Masked Prediction of Hidden Units},
year = {2021},
issue_date = {2021},
publisher = {IEEE Press},
volume = {29},
issn = {2329-9290},
url = {https://doi.org/10.1109/TASLP.2021.3122291},
doi = {10.1109/TASLP.2021.3122291},
journal = {IEEE/ACM Trans. Audio, Speech and Lang. Proc.},
month = oct,
pages = {3451–3460},
numpages = {10}
}

@inproceedings{Chen2024LDM-SVC,
  title     = {LDM-SVC: Latent Diffusion Model Based Zero-Shot Any-to-Any Singing Voice Conversion with Singer Guidance},
  author    = {Shihao Chen and Yu Gu and Jie Zhang and Na Li and Rilin Chen and Liping Chen and Lirong Dai},
  booktitle = {Proceedings of Interspeech 2024},
  pages     = {2770, 2774},      
  year      = {2024},
  address   = {Incheon, Korea}, 
  publisher = {ISCA},
  url       = {https://www.isca-archive.org/interspeech_2024/chen24e_interspeech.html}
}

@inproceedings{DDPM,
author = {Ho, Jonathan and Jain, Ajay and Abbeel, Pieter},
title = {Denoising diffusion probabilistic models},
year = {2020},
isbn = {9781713829546},
publisher = {Curran Associates Inc.},
address = {Red Hook, NY, USA},
booktitle = {Proceedings of the 34th International Conference on Neural Information Processing Systems},
articleno = {574},
numpages = {12},
location = {Vancouver, BC, Canada},
series = {NIPS '20}
}

@INPROCEEDINGS{DiffWaveNetSVC,
  author={Zhang, Xueyao and Fang, Zihao and Gu, Yicheng and Chen, Haopeng and Zou, Lexiao and Zhang, Junan and Xue, Liumeng and Wu, Zhizheng},
  booktitle={2024 IEEE Spoken Language Technology Workshop (SLT)}, 
  title={Leveraging Diverse Semantic-Based Audio Pretrained Models for Singing Voice Conversion}, 
  pages={758-765},
  doi={10.1109/SLT61566.2024.10832319}}

@inproceedings{crepe,
author = {Kim, Jong Wook and Salamon, Justin and Li, Peter and Bello, Juan Pablo},
title = {Crepe: A Convolutional Representation for Pitch Estimation},
year = {2018},
publisher = {IEEE Press},
url = {https://doi.org/10.1109/ICASSP.2018.8461329},
doi = {10.1109/ICASSP.2018.8461329},
booktitle = {2018 IEEE International Conference on Acoustics, Speech and Signal Processing (ICASSP)},
pages = {161–165},
numpages = {5},
location = {Calgary, AB, Canada}
}

@inproceedings{KID,
  author       = {Mikolaj Binkowski and
                  Danica J. Sutherland and
                  Michael Arbel and
                  Arthur Gretton},
  title        = {Demystifying {MMD} GANs},
  booktitle    = {6th International Conference on Learning Representations, {ICLR} 2018,
                  Vancouver, BC, Canada, April 30 - May 3, 2018, Conference Track Proceedings},
  publisher    = {OpenReview.net},
  year         = {2018},
  url          = {https://openreview.net/forum?id=r1lUOzWCW},
  bibsource    = {dblp computer science bibliography, https://dblp.org}
}

@article{MMD,
  title        = {A Kernel Two-Sample Test},
  author       = {Arthur Gretton and Karsten M. Borgwardt and Malte J. Rasch and Bernhard Sch{\"o}lkopf and Alexander Smola},
  journal      = {Journal of Machine Learning Research},
  volume       = {13},
  number       = {25},
  pages        = {723--773},
  year         = {2012},
  url          = {http://jmlr.csail.mit.edu/papers/v13/gretton12a.html}
}
\endgroup

\end{document}